\documentstyle[twocolumn,aps]{revtex}
\begin{document}
\draft
\twocolumn[\hsize\textwidth\columnwidth\hsize\csname @twocolumnfalse\endcsname
%
%
%

\title{Spin Dynamics of Hole Doped
${\rm Y_{2-x} Ca_x Ba Ni O_5}$}

\author{ E. Dagotto$^1$,   J. Riera$^2$, A. Sandvik$^1$, and A. Moreo$^1$}

\address{$^1$Department of Physics and National High Magnetic Field Lab,
Florida State University, Tallahassee, FL 32306, USA
\\
$^2$ Instituto de F\'isica Rosario y Depto. de F\'isica, Univ. Nac. de
Rosario, 2000 Rosario, Argentina}

\date{\today}
\maketitle

\begin{abstract}
We propose an electronic model for the recently discovered
hole doped compound ${\rm Y_{2-x} Ca_x Ba Ni
O_5 }$. From a multiband Hamiltonian with oxygen and nickel
orbitals, a one band model is derived.
Holes are described using Zhang-Rice-like
S=1/2 states at the nickels  propagating on a S=1 spin chain.
Using numerical techniques to calculate the dynamical spin structure
factor ${\rm S(q,\omega)}$ in a realistic regime of couplings, spectral weight
in the Haldane gap is observed in agreement with
neutron scattering data.  Low energy
states with S=3/2 appear in the model. Several predictions are made to
test these ideas.

\end{abstract}

\vskip2pc]
\narrowtext

%
%

Spin-liquid ground states in
Heisenberg S=1 chains and S=1/2 ladders
have been studied theoretically as paradigms
of disordered nonclassical systems.\cite{haldane,ladder}
These spin models can be physically
realized in several compounds.
An important issue is the effect of hole doping on these
systems. Theoretical studies of doped S=1/2 ladders have shown that
the spin gap survives in the presence of holes, and that the spin-gapped
phase is favorable for superconductivity.\cite{ladder} Behavior
indicating a spin gap has also been observed experimentally in some
underdoped high-Tc cuprates. However, the case of the doped S=1 chains
has been considered only very recently
in the context of the S=1 metal oxide
${\rm Y_{2} Ba Ni O_5 }$.\cite{batlogg}
Lightly doping this compound with Ca introduces
hole carriers in the chains.
Two remarkable experimental features were observed upon doping, namely  (i)
the reduction of
the resistivity $\rho_{dc}$ by several orders of magnitude
leading to a one dimensional conductor, and (ii)
the creation of states inside the Haldane gap
as revealed by inelastic neutron scattering (INS)
data.\cite{ditusa}

To understand these in-gap states, $spin$ systems
with site or bond impurities have been recently proposed.\cite{yulu}
Holes  are assumed to be so strongly localized that
their mobility is neglected. In another approach dynamics
was provided to the holes\cite{shiba} but the Ni-Ni exchange J was
assumed larger than both the hole hopping amplitude and
the short-bond NiO exchange.
While studying these limits is instructive since gap states
are produced, the drastic reduction in $\rho_{dc}$ observed experimentally
upon doping\cite{ditusa}
suggests
that holes are likely mobile
over several lattice spacings.\cite{comm5}
Thus, in this paper we propose a new model for ${\rm Y_{2-x} Ca_{x} Ba Ni O_5}$
with fully $mobile$ S=1/2 holes interacting with S=1 spins, which is
studied using realistic values for the couplings.

In ${\rm Ni^{++}}$ surrounded by oxygens,
${\rm d_{3z^2 - r^2}}$ and ${\rm d_{x^2 - y^2}}$ are the two active
orbitals if deviations from the perfect
octahedral symmetry in ${\rm NiO_6}$ are considered.\cite{comm6}
Then, as a Hamiltonian for the ${\rm Ni-O}$ chains in the hole notation
we propose
$$
{\rm H =
- \sum_{\langle ij \rangle, \sigma, \alpha} t_{pd \alpha}
(d^{\dagger}_{i \sigma \alpha} p_{j \sigma} + h.c.) +
U_d \sum_{i,\alpha} n^{\alpha}_{i\uparrow} n^{\alpha}_{i\downarrow} }
$$
$$
{\rm +
U_p \sum_{j} n_{j\uparrow} n_{j\downarrow} +
\Delta \sum_{j} n_j - |J_{Hund}| \sum_{i} {\bf S_{i1}}.{\bf S_{i2}}}.
\eqno(1)
$$
\noindent ${\rm i}$ (${\rm j}$) denotes ${\rm Ni}$ (${\rm O}$)
sites. $\alpha = 1 (2)$ corresponds to the Ni orbitals ${\rm d_{3z^2 -
r^2}}$ (${\rm d_{x^2 - y^2}}$).
The Coulomb repulsion is ${\rm U_d}$ (${\rm U_p}$) at
the ${\rm Ni}$ (${\rm O}$) sites,  and $\Delta$ is the
charge-transfer
energy. ${\rm d_{i \sigma \alpha}}$ are $hole$ operators
corresponding to a ${\rm Ni}$ site, spin $\sigma$ and orbital $\alpha$,
while ${\rm p_{j \sigma}}$ are oxygen hole operators.
The last term in Eq.(1) is a $ferromagnetic$
coupling between the Ni holes on different
orbitals
(using ${\rm {\bf
S}_{{\bf i} \alpha} =  d^\dagger_{{\bf i} \alpha} {\bf \sigma}
d_{{\bf i} \alpha}/2 }$), which enforces
Hund's rule. This term is important to produce the expected S=1 state
in ${\rm Ni^{++}}$.
{}From an analysis of various
spectroscopies for ${\rm Li_{x} Ni_{1-x} O}$, it was found
that ${\rm \Delta = 6.0 eV}$, ${\rm U_{d} = 9.5 eV}$, and
${\rm U_{p} = 4.6 eV}$.\cite{kuiper}
Since ${\rm U_{d} > \Delta}$ then
the compound is in the charge-transfer regime.\cite{charge}
${\rm J_{Hund}}$ is obtained from the energy
levels of a
${\rm {(Ni O_6)}^{10-} }$ cluster.\cite{lines} The difference in energy (${\rm
|J_{Hund}|}$)
between the S=1 $^3A$ and the S=0 $^1E$ levels is 1.3eV.
The hopping amplitudes
are ${\rm t_{pd1} = 1.3 eV}$ and
${\rm t_{pd2} = 0.75}$, according to a cluster
calculation.\cite{vanelp,harrison}

The study of mobile holes in model Eq.(1)
is a difficult task
since the energy scale is eV, while the interesting physics
for ${\rm Y_2 Ba Ni O_5}$ occurs at the J-scale of about
30 meV. Thus, we need to construct an effective low-energy
Hamiltonian from Eq.(1) which here we carry out following the work of
Zhang and Rice
in their reduction of the 2D multiband Hubbard model
to the t-J model.\cite{zhang} In
${\rm Y_{2-x} Ca_x Ba Ni O_5}$
holes populate oxygens,\cite{ditusa}
and they have a strong exchange ${\rm J'}$ with neighboring Ni.
This may lead to low energy states with S=1/2 or
S=3/2,  after the oxygen S=1/2 hole mixes with neighboring S=1 spins.
To study this effect,
we solved a small cluster O-Ni-O described by Eq.(1) in the three  holes
subspace (two of them producing the Ni S=1, and the other
providing an extra oxygen hole). From the spectrum, and using the parameters
given before,
we indeed found a S=1/2 ground state, with the
S=3/2 state  located ${\rm \sim 1.3}$ eV higher.\cite{zhang2}
This energy difference
is robust even if
the parameters of Eq.(1) are modified within acceptable windows
to account for experimental uncertainties.
Then, proceeding \'a la Zhang-Rice we project out
all the states of the three holes
O-Ni-O cluster but the  Ni-centered S=1/2 ground state.\cite{comm3}
Eq.(1) thus reduced to an effective model with
states $only$ at the Ni sites  which can be S=1 (``spins'')
or S=1/2 (``holes''). Thus, in our model
the study of carriers on ${\rm Y_{2-x} Ca_x Ba Ni
O_5 }$ amounts to the analysis of
S=1/2 hole-like states
in a S=1 background.\cite{comm}

To derive the allowed hole hopping
processes it is important to remember the composite character
of the S=1/2 hole state. To guide the intuition a graphical
representation is useful (Fig.1a). The low energy
states with S=1 and 1/2 located
on nearest neighbors Ni sites
are actually represented as five holes (four in the Ni d-shells  and
one in the O p-shell) on a Ni-O-Ni cluster. The dashed line
signals a possible singlet between
the O-hole with one of the Ni-holes producing a S=1/2 state as was
described before.
However, it is clear that the
Ni partner of the oxygen hole can be easily switched from right to
left producing an effective hopping process of the S=1/2
low energy state as shown in
Fig.1a. Other hoppings do not correspond to
a mere interchange of the states but a spin flip can also occur.
Exploring all possibilities, it can be easily shown that
the allowed processes  are those listed
in Fig.1b (plus their spin reversed analogs). These hopping terms
have been also derived by Zaanen and Ole\'s\cite{zaanen2}
using more formal arguments
in their complementary analysis of triplet holes moving in a S=1/2 background.

The hopping processes (Fig.1b) can be written in a compact, explicitly
rotationally invariant form. The full one-band Hamiltonian then becomes
$$
{\rm H = J \sum_{\langle ij \rangle} {{\bf S}_{ i}}\cdot{{\bf S}_{j}} - t
\sum_{\langle ij \rangle} P_{ij} ( {\bf \hat{S}}_{i}\cdot{\bf \hat{S}}_{j} +
1/2 ) }.
\eqno(2)
$$
\noindent The first term is the Heisenberg interaction
 arising from Eq.(1) at half-filling and strong coupling. It only affects
the S=1 spins. Calculating J directly from Eq.(1) is
difficult and thus here we simply take the
exchange from experiments i.e. ${\rm J = 0.03 eV}$. The second term in
Eq.(2) acts only when spins 1 and 1/2
share a link. ${\rm P_{ij} }$
simply permutes the two spins. The notation
${\rm {\bf \hat{S}}_{i} = ( \hat{S}_x,\hat{S}_y,\hat{S}_z ) }$  is used
in the hopping term to indicate that
this spin operator can act over both S=1 $and$ 1/2 states depending
on what spin is located at site ${\rm i}$. Their action is the standard
for a spin operator. For example,
${\rm { \hat{S}}^-_{i} { \hat{S}}^+_{j}
| m^z_{i} = 1; m^z_{j} = -1/2 \rangle = \sqrt{2}| m^z_{i} = 0; m^z_{j} = +1/2
\rangle }$. Finally, it is natural to assume that
the realistic coupling regime is
${\rm t > J}$ although we have not derived t and J from Eq.(1).\cite{comm30}

An accurate analysis of one dimensional
(1D) Hamiltonians like Eq.(2) can be done using exact diagonalization (ED)
techniques.\cite{review,monte}
With this method, we first calculated
the dispersion of $one$ hole, ${\rm \epsilon(q)}$, using model Eq.(2) (see
Fig.2a).
${\rm \epsilon(q)}$
is not symmetric with respect to
$\pi/2$ contrary to systems with antiferromagnetic order.
The
minimum ${\rm q_{min}}$ at ${\rm t/J = 2.0}$ is at $\pi/3$ for the chain
used in Fig.2a. In
the bulk limit we expect that
${\rm q_{min}}$ will move with continuity as a function
of ${\rm t/J}$.
While the hole ground state at  most q values has S=1/2,
the states close to ${\rm q = \pi}$ have S=3/2
i.e. the band in Fig.2a is actually made out of two energy levels with
different spins.
In the range ${\rm 0.5 \leq t/J  \leq 4.0}$, the bandwidth W is approximately
${\rm 0.2-0.3 J}$  (Fig.2b)i.e. much smaller than the ${\rm W \sim
2J}$  reported for
holes in 2D S=1/2 antiferromagnets.\cite{review}
Several of these features
could be observed in angle-resolved photoemission experiments.
The peculiar dynamics of Eq.(2) is responsible for the abnormally small W.
Such a large effective
mass increases the tendency to hole localization,  providing a
natural explanation for  the
lack of true metallicity in ${\rm Y_{2-x} Ca_x Ba Ni O_5 }$
and  in doped 2D nickelates.

Fig.2c shows
${\rm S( \pi,\omega)}$ at a nominal hole density
${\rm x = 1/12 \sim 0.083}$ and ${\rm t/J = 2.0}$
calculated with ED.\cite{comm12}
The qualitative agreement
with the INS data reported for
${\rm Y_{2-x} Ca_x Ba Ni O_5 }$\cite{ditusa} at ${\rm x \sim 0.04 - 0.10}$ is
clear. The spectrum
shows two main features: peak I caused by spin excitations
in the S=1 chain away from the hole, and peak II induced by the
addition of a S=1/2 hole.
To justify these identifications, let us study the q-dependence of
peaks I and II (Fig.3a).
Since ${\rm S(q, \omega)}$ at finite x
has low-intensity structure in addition to the
two main peaks (see inset
of Fig.2c), the results shown in
Fig.3a are just
rough estimations but we believe the main qualitative features are
robust. Clearly peak I follows closely the position of the one magnon
excitation observed in undoped S=1 chains, while
peak II is virtually
dispersionless.
Both peaks rapidly loose intensity
when moving away from $\pi$, in agreement with INS \cite{ditusa} results
where only ${\rm q=\pi}$ provided a large signal.\cite{comm11}

To better understand peak II let us study ${\rm
S(q,\omega) }$ for the undoped S=1 chain with $open$ boundary
conditions (OBC) (Fig.3b). At the ends of the chain, extended S=1/2
states are formed\cite{extension} which are coupled forming singlets (S) and
triplets (T),  becoming degenerate as ${\rm N \rightarrow
\infty}$. Peak II in Fig.3b corresponds to the
transition between the S and
T states. We believe this transition also
contributes to ${\rm S(\pi, \omega) }$ in Fig.2c.
We explicitly checked that the main
contribution to peak II in Figs.2c and 3b
comes from spins located near the $ends$ of the
S=1 chain. Note that on Fig.3b
it seems strange that the strength of peak
II is relatively large since
it should vanish in the bulk recovering
the Haldane gap ${\rm \sim 0.4J}$ (peak I). The reason is
that the ``size'' of the S=1/2 end-states is
$\sim 6$ lattice spacings\cite{extension} and thus
only when ${\rm N \gg 12}$ (or the density of hole defects is
${\rm x \ll 1/12}$)
is that the contribution of the S-T transition will be negligible. This
lead us to predict that for $zinc$ doping ${\rm y \sim 0.10}$,
substantial weight inside the Haldane gap should be observed with
neutron scattering as for the case of Ca doping
(experiments thus far have only been done for ${\rm y =
0.04}$ \cite{ditusa}).

To understand the interaction between the S and T states with the hole
consider first the unphysical limit ${\rm t\ll J}$.
In this case at low energies the
dominant states are formed by the direct product of the S and T with the hole
S=1/2. For a ring with N finite and $even$,
the end-states are in a triplet state  separated by a gap from
the singlet combination. The triplet T couples to the S=1/2 hole
forming states of S=1/2 and 3/2, denoted here as ${\rm | 1/2(T) \rangle}$ and
$| 3/2 \rangle$, respectively.
These states  are split by a finite ${\rm t/J}$, and
${\rm | 1/2(T) \rangle}$  becomes
the lowest energy state for a dynamical hole.
The singlet S
forms a S=1/2 state with the hole (${\rm | 1/2(S) \rangle}$ )
which is located at a higher
energy.
In Fig.3c results are shown for ${\rm | 1/2(T) \rangle}$
and $| 3/2 \rangle$ at ${\rm t/J= 0.1}$.
If ${\rm N}$
is $odd$ the positions of the quasi-degenerate sextuplet
${\rm | 1/2(T) \rangle }$ -- ${\rm | 3/2 \rangle}$, and the doublet
${\rm | 1/2(S) \rangle }$ are reversed.
It is remarkable that
increasing ${\rm t/J}$ to realistic values
the energy difference between these states remain a fraction of J.
For example, in Fig.3c  the energies of the  ${\rm | 1/2(T) \rangle}$
and ${\rm | 3/2\rangle}$
states are also shown at
${\rm t/J = 2.0}$. The low energy physics in
${\rm S(q, \omega) }$ (at ${\rm x > 1/12}$) is still
dominated by the transitions between these two states.
If x is reduced from ${\rm \sim 0.08}$ the
relative importance of these states should rapidly decrease, but for the
densities used in INS (${\rm x = 0.10}$)
they are very important as already shown in Fig.2c.

Finally note that as ${\rm t/J}$ increases further, the kinetic energy becomes
dominant and the
hole prefers to be surrounded by a cloud of
ferromagnetically aligned spins. On the N=12 chain, the hole state with
S=3/2 becomes the ground state at ${\rm t/J \sim 5.74}$. This coupling is
not unrealistic and thus the high-spin states observed in the
magnetic susceptibility\cite{batlogg} could be
caused by the presence of low energy
S=3/2 Zhang-Rice holes.\cite{comm8}
We also found that studying $two$ holes on a 10 sites
chain (density ${\rm x = 2/10
\sim 0.20}$),
the remnants of the Haldane gap are difficult
to identify in ${\rm S(q, \omega)}$.
This prediction can be tested experimentally increasing x in
${\rm Y_{2-x} Ca_x Ba Ni O_5 }$ beyond the current limit ${\rm x=0.10}$.

Summarizing, a model for $mobile$ carriers in NiO chains was here
proposed.
${\rm S(q,\omega)}$ shows Haldane gap states in
agreement with neutron scattering data.
We believe that theories where holes
are modeled as spin impurities can be distinguished from the
dynamical holes approach described here by measuring the optical
conductivity. The present theory would
predict a substantial weight
inside the charge transfer gap upon doping since holes are highly mobile
and finite $\omega$ precursors of Drude peaks should appear in severed
NiO chains.

We thank G. Zimanyi, J. Zaanen, A. Fujimori, J. DiTusa, G. Aeppli,
N. Bonesteel, and B. Batlogg for useful conversations.
This work was supported by ONR under
N00014-93-0495. We also thank Martech for support.
\medskip

\vfil\eject

%
%

{\bf Figure Captions}

\begin{enumerate}

\item (a) Graphical representation of a hopping process as explained in
the text; (b) Allowed hopping processes in
model Eq.(2). The numbers in the ket represent the z-projection
of the S=1 and 1/2 spin states of Eq.(2) at two arbitrary sites i and j,
respectively.
${\rm P_{ij} 1/2}$ in Eq.(2) enforces the $absence$ of the hopping
${\rm |1,-1/2 \rangle \rightarrow |-1/2,1 \rangle}$.
The spin projections $only$ change by 1/2 or -1/2 at
each site.

\item
(a) ${\rm \epsilon(q)}$ on a ${\rm N=12}$ chain with one hole and
periodic boundary
conditions (PBC) using Eq.(2). The energy is measured with respect
to the lowest energy state. Open (full)
circles correspond to ${\rm t/J = 0.5}$ ($2.0$):
(b) Hole bandwidth W (energy difference between the highest
and lowest energy in ${\rm \epsilon(q)}$) vs ${\rm t/J}$ on the N=12
chain;
(c) ${\rm S(\pi,\omega)}$ of model Eq.(2) with
${\rm t/J=2.0}$, ${\rm N=12}$, one hole and
PBC. The meaning of peaks I and II is discussed in the text. In the
inset ${\rm S(\pi/2,\omega)}$ is shown for the same set of parameters.

\item
(a) Energy of the excitation with the highest intensity in ${\rm
S(q,\omega)}$ (parameters as in Fig.2c). The
full squares  denote peaks I and II as they evolve with
q, with the error bars representing their width. The open squares are
the one magnon peak position on a
${\rm N=14}$ undoped S=1 Heisenberg model with PBC;
(b) ${\rm S(\pi,\omega)}$ for the undoped S=1 Heisenberg model on
a 12 sites chain with OBC. The meaning of the peaks is discussed in the text;
(c) Energy vs q of the relevant low energy states (8 sites chain with
PBC and one hole). Open squares and circles  denote
the states (defined in the  text)
${\rm |1/2(T) \rangle }$ and ${\rm | 3/2
\rangle }$,respectively, at ${\rm t/J = 0.2}$. Full
squares and circles
correspond to
${\rm |1/2(T) \rangle }$ and ${\rm | 3/2
\rangle }$, respectively,
at ${\rm t/J=2.0}$.

\end{enumerate}


\begin{references}
\bibitem{comm5} We believe the entire compound does not
behave like a pure metal
because in 1D a small concentration of defects
can strongly influence the d.c. conductivity.
\bibitem{comm6} The
Jahn-Teller induced splitting will be neglected here.
\bibitem{comm30} A
Heisenberg interaction ${\rm J_{1-1/2}}$
(${\rm J_{1/2-1/2}}$)  between one S=1 and a hole
(two holes) is here neglected. For realistic couplings, t is
more relevant for the hole properties than ${\rm J_{1-1/2}}$ and
${\rm J_{1/2-1/2}}$.
\bibitem{monte} We have found that Monte Carlo techniques for
 Eq.(2) have sign problems even in 1D.
\bibitem{comm12} Note that the one hole ground state has a finite
momentum, as shown in Fig.2a, which complicated the analysis of
${\rm S(q, \omega)}$. However, we believe the main results shown in this paper
are not affected by this detail. A study of 2 holes on a 24 site chain
is not possible numerically.
\bibitem{comm11} While peak II in Fig.2c seems sharply located at about half
the Haldane
gap, the experimental situation may correspond to a random average of
severed chains with different lengths including ${\rm N < 12}$
where the energy of peak II is modified by size effects. This smears
the low energy peak improving the agreement with experiments.
\bibitem{comm8} This state should $not$ be confused with the much higher
energy S=3/2 state that appeared in
the O-Ni-O cluster during the Zhang-Rice state construction (Ref.17).
\end{references}
\end{document}